# Quantifying Device Usefulness - How Useful is an Obsolete Device?

Craig Goodwin[1] [0000-0003-3296-7200], Sandra Woolley[1] [0000-0002-7623-2866], Ed de Quincey[1][0000-0002-3824-4444] and Tim Collins[2][0000-0003-2841-1947]

[1] Keele University, Staffordshire, UK

[2] Manchester Metropolitan University, Manchester, UK

c.goodwin@keele.ac.uk

**Abstract.** Obsolete devices add to the rising levels of electronic waste, a major environmental concern, and a contributing factor to climate change. In recent years, device manufacturers have established environmental commitments and launched initiatives such as supporting the recycling of obsolete devices by making more ways available for consumers to safely dispose of their old devices. However, little support is available for individuals who want to continue using legacy or 'end-of-life' devices and few studies have explored the usefulness of these older devices, the barriers to their continued use and the associated user experiences. With a human-computer interaction lens, this paper reflects on device usefulness as a function of utility and usability, and on the barriers to continued device use and app installation. Additionally, the paper contributes insights from a sequel study that extends on prior work evaluating app functionality of a 'vintage' Apple device with new empirical data on app downloadability and functionality for the same device when newly classified as 'obsolete'. A total of 230 apps, comprising the top 10 free App Store apps for each of 23 categories, were assessed for downloadability and functionality on an Apple iPad Mini tablet. Although only 20 apps (8.7%) could be downloaded directly onto the newly obsolete device, 143 apps (62.2%) could be downloaded with the use of a different non-legacy device. Of these 163 downloadable apps, 131 apps (comprising 57% of all 230 apps and 80.4% of the downloadable apps) successfully installed, opened, and functioned. This was a decrease of only 4.3% in functional apps (of the 230 total apps) compared to the performance of the device when previously classified as 'vintage'.

## 1   Introduction

Sustainable HCI [6, 14, 27, 29] and the study of device longevity and usefulness are particularly important whilst the number of obsolete devices and the levels of global e-waste continue to rise. The "Internet of Trash" [7, 15] has been used to describe the billions of end-of-life mobile and Internet-connected devices [9, 18] that contribute to the 53 million tons of e-waste generated per year [10]. In a review of the literature,

Mellal [22] compares and contrasts different definitions of 'obsolete' and 'obsolescence' and distinguishes between types of obsolescence such as 'technological', 'functional', 'style' and 'planned'. Planned obsolescence is a contentious issue [4, 8, 20]. Though it has been argued to be a consequence of competitive forces in a free and technological society [32] it contributes to increasing sustainability concerns and to consumer dissatisfaction [19, 21, 30] particularly amongst users of not-so-new devices. However, it is not unusual for device manufacturers to launch new device models and variants on an annual basis [31].

Apple has made a commitment to carbon neutrality by 2030 [2] and has committed to improving product recycling and the use of recycled materials. Currently, however, little is known, in general, about where obsolete products go after being sent for recycling [13]. In terms of definitions, Apple defines products as 'vintage' when "Apple stopped distributing them for sale more than 5 and less than 7 years ago" and defines them as 'obsolete' when "Apple stopped distributing them for sale more than 7 years ago" [3]. When devices are 'vintage' (but not 'obsolete') they are in a transitional state, where support from app developers declines, updates reduce, and users may receive warnings that apps will no longer be supported [1]. When devices become 'obsolete' their warranties expire and the services that Apple were legally obliged to provide previously will no longer be available [3].

Often obsolete or 'end-of-life' devices hold little or no value in terms of serving their original purpose [26, 34] and, to date, few studies have focused on the assessment of the usefulness of vintage or obsolete devices. Where legacy devices are reused, their applications are often limited in scope (at least compared to their original lives as more general-purpose computing devices) and can be trivial compared to their original capability. For example, a legacy iPad used as a shopping list, or an iPhone used as a music player [28]. While this sort of repurposing extends the lifespan of devices and delays their disposal, if we think about usefulness with an HCI lens as a function of utility and usability [16, 25] then the limited nature of this type of repurposed utility inevitably reduces the device usefulness [5]. So how useful can a legacy device be? Can we quantify its usefulness? If utility relates to the scope of device use, then the ability to continue installing and updating apps must be significant to its usefulness. In this paper we investigate device usefulness by exploring the barriers to software installation and analyzing the functionality of downloadable apps. The work extends on a prior study that evaluated app functionality of a 'vintage' device [12] and it contributes new empirical results that quantify app downloadability and functionality for the same device when newly 'obsolete'. The study results are compared, and the usefulness and user experiences of vintage and obsolete devices are reflected upon.

## 2   Methods

Attempts were made to install popular free Apple App Store apps onto an obsolete device. The study device was an Apple iPad Mini Tablet, first manufactured in 2012, discontinued in 2015, received the last OS update (iOS 9.3.5) in 2016 and classed as 'obsolete' by Apple in 2022 [3]. The study took place in a three-day period between 4th and 6th May 2022. This allowed just enough time to attempt downloads for all top 10 free apps from 23 App Store categories and minimized the risk of apps having a newer pushed update, meaning the last supported versions could be removed from the Apple App Store.

For apps that cannot be downloaded directly, a current non-legacy Apple device must be used to obtain a "purchase history" on an Apple account that is shared with the legacy device. An attempt can then be made to download the "last previously supported" app for the obsolete device. Alternatively, users can connect their device to a computer and use an older version of Apple's "iTunes" to download the required app. But, either way, another device is necessary for the app installation process and these methods are not well-known and feature only infrequently on Apple forums [23]. For simplicity, we refer to this somewhat complicated workaround to downloading as 'Download via Another Device' (DvAD). To obtain a purchase history on a shared Apple account, a current non-vintage iPhone SE was used.

### 2.1 App Selection Criteria

A total of 230 apps were selected comprising the top 10 free App Store apps from 23 categories. App categories requiring modern features such as AR mode and extensions for the Apple Watch were excluded due to their incompatibility with the study device.

As shown in Fig. 1, each of the top ten apps for each of the categories was tested to determine whether it could be downloaded directly. Apps that did download were tested to determine whether they installed, opened, and functioned. Attempts were made to download apps via another device (as summarized earlier) if they did not download directly. Apps that downloaded successfully in this way were then tested to determine whether they installed, opened, and functioned.

## 3 Results

As shown in Fig. 1, only 20 (8.7%) of the total 230 top apps could be directly downloaded and of these, only 16 apps (7%) functioned. However, 143 apps (62.2%) of the total 230 top free apps could be downloaded with the help of another device, making a total of 163 apps (70.1% of all apps) that could be downloaded either directly or via another device. Of the 163 apps that did download, 115 (80.4% of the 163 apps) installed, opened, and functioned.

In total, 131 out of the 230 (57%) total apps could either be downloaded directly or via another device and were capable of functioning. This was a decrease of only 4.3% in functional apps (of the 230 total apps) compared to the performance of the device when previously classified as 'vintage'. In total, 67 (29.1% of the 230 apps) were not downloadable, however, 27 (40.3% of the 67 non-downloadable apps) were never previously supported by the device. For example, these included apps that had a release date long after the device was originally released. In total, 99 out of 230 (43%) apps were unsuccessful in download, installation, opening and/or functioning. However, of the 163 apps that did download, 131 apps (80.4%) successfully installed, opened, and functioned.

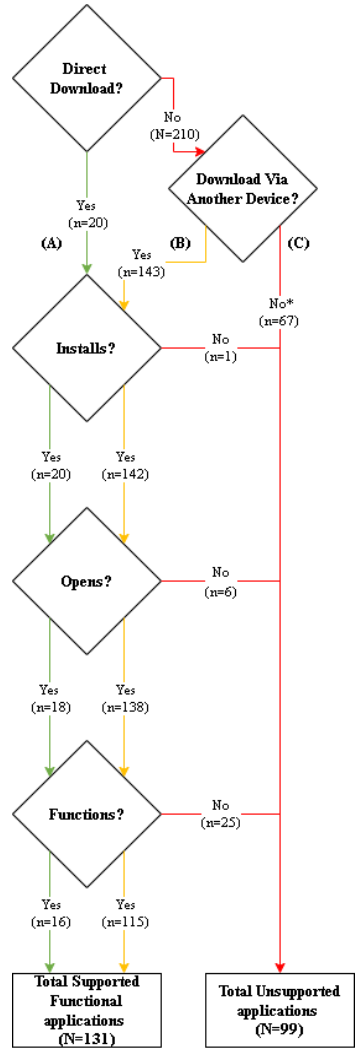

**Fig. 1.** A summary of the study method and app results. (A): Apps downloaded directly (green). (B): Apps downloaded via another device (yellow). (C): Apps that could not be downloaded directly or via another device (*including the 27 apps were never previously supported by the device iOS version) or that failed "Installs?", "Opens?" or "Functions?" (red).

## 3.1   Result Breakdown by Category

A breakdown of functional apps by category is shown in Fig. 2. Although dominated by apps that required download via another device, at least half of the apps in 18 of the 23 categories successfully functioned and more than half (i.e., at least 6 out of 10) apps successfully functioned for 14 of the 23 categories. Of the apps that could be downloaded, six failed to open and 25 failed to function but only one (Google maps in the 'Navigation' category) failed to install.

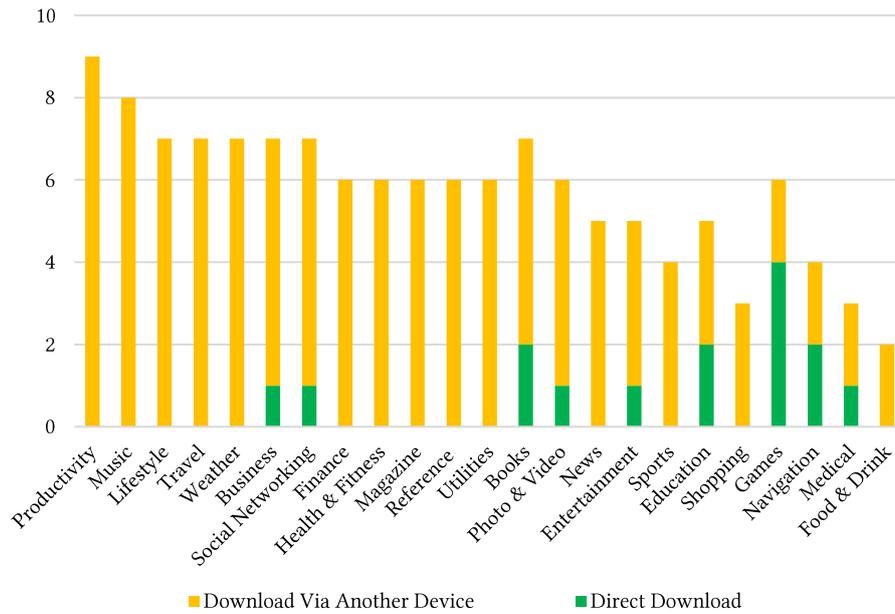

**Fig. 2.** Functional Apps by App Category

## 4  Analysis and Discussion

Overall, the study demonstrated that the majority of apps installed, opened and functioned on the obsolete device if they could be downloaded, and that only a small reduction in functioning apps occurred between the vintage and obsolete studies.

Apps were considered 'functional' if they performed key functions as intended. For example, if a video streaming app could play a video, or if a game was playable. However, it was not possible to confidently assess the functionality of all apps. For example, some apps in Finance, Food & Drink and Utilities categories require logins to pre-existing user accounts to unlock features, make purchases and manage accounts. In these cases, minimal functionality was assumed if the apps opened with a login screen with no warning or incompatibly notification. For example, some apps in the Entertainment category successfully downloaded, installed, and opened but with a notification that the app was no longer compatible and an upgrade was recommended.

### 4.1 Comparing 'Vintage' and 'Obsolete' Device Results

A summarized performance comparison of the two studies is provided in Table 1. The number of apps that could be downloaded directly reduced from 29 apps (12.6% of the top 230 apps) in Sept 2021 when the device was vintage down to 20 apps (8.7% of the top 230 apps) when the device was newly obsolete. Also, the number of apps that could be downloaded (either directly or via another device) increased slightly from 140 apps (60.9%) to 143 apps (62.2%). However, the number of apps that functioned (having been downloaded by either method) reduced from 141 apps (61.3%) when the device was vintage to 131 apps (57%) when the device was newly obsolete. This small decrease in functional apps might be expected as a device enters its obsolete phase.

Between the two studies there were some variations in the contemporary rankings and memberships of the free top 10 apps, and there were some variations in the performances of apps in each category. However, in both studies, all top 10 'Productivity' category of apps (e.g., email and calendar apps) failed to download directly yet 9 of the 10 apps functioned on the obsolete device and all 10 functioned on the vintage device after successful download via another device. Similarly, for the 'Health and Fitness' and 'Travel' categories of apps there were no changes in performance between the two studies yet 6 out of the 10 apps on the vintage device and 7 out of 10 apps on the obsolete device, functioned after successful download via another device. In contrast the number of functional 'Games' apps increased by 5 from 2 to 7 on the obsolete device (four being directly downloadable) which was not anticipated due to i) the release dates of popular games being rather more recent than the device, and ii) the minimum hardware requirements of games might often be expected to exceed those of a legacy device.

**Table 1.** Comparison of the results from each study

|  | Sept 2021 | May 2022 | Percentage Change |
|---|---|---|---|
| Direct Download | 12.6% | 8.7% | -3.9% |
| Download via Another Device | 60.9% | 62.2% | +1.3% |
| Non-Downloadable | 26.5% | 29.1% | +2.6% |
| Direct Download and Functions | 10.4% | 7.0% | -3.4% |
| Download via Another Device and Functions | 50.9% | 50% | -0.9% |
| Total Functional Apps | 61.3% | 57% | -4.3% |
| Total Non-functional* Apps | 38.7% | 43% | +4.3% |

*'non-functional' apps are apps that did not download (either directly or indirectly) or did not install, open or function.

### 4.2 The Usefulness of an 'Obsolete' Device

Labels like "vintage" and "obsolete" may make consumers perceive devices as no longer functional, usable or useful and, therefore, ready for disposal. However, as the study results demonstrate, this is not the case. Legacy devices, whether 'vintage' or 'obsolete' devices, can be capable of extended and useful function that includes the installation of many new apps. But at what point do consumers give up on efforts to continue making use of their devices? Perhaps another decrease in functioning apps at this boundary between vintage and obsolete classifications is the point at which all but the most determined users with the necessary app download know-how and access to a non-vintage device, will give up on new app installations and further use of their device.

Ideally, manufacturers would promote device longevity and assist consumers in avoiding barriers to extending the lifespans and usefulness of their devices [17, 24]. As illustrated in the percentage of functional apps in Fig. 3, if we consider that usefulness correlates with apps capable of functioning (not only via direct download but by either method) then vintage and obsolete devices clearly have potential for longer, useful lifespans.

We cannot forget, however, that security is a key consideration for all devices. It could be argued that obsolete devices are not viable platforms given current challenges to safe and secure systems. Certainly, apps where security is critical, for example, banking apps, should not be used on obsolete devices. However, many useful apps like games, calculators, media players and other tools may present relatively little, or no, security concerns.

Graceful degradation could be a method that manufacturers adopt in future iterations of their device ranges as a way of being more inclusive to legacy device users [33]. This would reduce barriers to device longevity and avoid sudden reductions in device usefulness. A gradual decline would likely be preferred by legacy device users but quantifying or being aware of this decline (see Fig. 3) is difficult for consumers to follow.

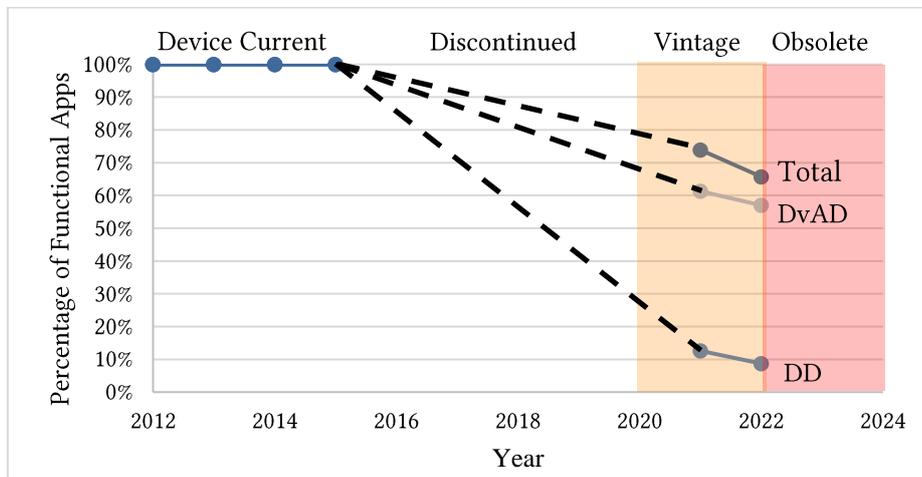

**Fig. 3.** Percentage of functional apps in the context of the device lifespan that could be downloaded directly (DD) or downloaded via another device (DvAD).

### 4.3 Study Limitations and Future Work

The study was limited to 230 popular free apps (10 from each of 23 categories). Increasing the number of apps and categories could provide more insights, however, there is an inherent time pressure to download apps because apps are updated regularly. Future work could extend the functionality assessments of apps and include assessments of their performance and usability which could include a more thorough evaluation of application functionality. Also, future work could apply a similar study methodology (as illustrated in Fig. 1) for the assessment of other Apple devices. As more devices become vintage and obsolete, a larger and more comprehensive comparative study could be conducted to assess and 'rank' changes in performance and usefulness across multiple systems. Furthermore, a similar study methodology could potentially be applied to other smartphone and tablet devices such as Android devices with "download via another device" replaced by the "sideloading" of apps [11]. This device compared to a similar era Android device has much less developer support and has a higher probability of becoming e-waste sooner as the limitations on usage become more prevalent.

## 5 Conclusion

The study demonstrated that most of the apps installed, opened, and functioned on the obsolete device if they could be downloaded, and that only a small reduction in functioning apps occurred between vintage and obsolete stages.

Continued efforts to improve product recycling must be made but, to reduce the mounting levels of electronic waste, new strategies are needed. As more devices become obsolete on an annual basis, new initiatives are needed to support users improve the longevity and usefulness of their old devices. It is recommended that device manufacturers remove the barriers to lifecycle extensions for obsolete devices by releasing patches to allow direct download of apps still compatible with legacy devices.

## References

1. Apple (2020) *Apple Lifecycle Management*, *Apple*. Available at: https://www.apple.com/ie/business/docs/resources/Apple_Lifecycle_Management.pdf (Accessed: January 11, 2023).
2. Apple (2021) *Apple charges forward to 2030 carbon neutral goal, adding 9 gigawatts of Clean Power and doubling supplier commitments*, *Apple Newsroom*. Available at: https://www.apple.com/newsroom/2021/10/apple-charges-forward-to-2030-carbon-neutral-goal-adding-9-gigawatts-of-clean-power-and-doubling-supplier-commitments/ (Accessed: January 11, 2023).
3. Apple (2023) *Obtaining service for your Apple product after an expired warranty*, *Apple Support*. Available at: https://support.apple.com/en-us/HT201624 (Accessed: January 12, 2023).
4. Barros, M. and Dimla, E. (2021) "From planned obsolescence to the circular economy in the smartphone industry: An evolution of strategies embodied in product features," *Proceedings of the Design Society*, 1, pp. 1607–1616. https://doi.org/10.1017/pds.2021.422.

5. Bieser, J. *et al.* (2021) *Lifetime extension of mobile internet-enabled devices: Measures, challenges and environmental implications*, *University of Zurich*. University of Limerick Institutional Repository. Available at: https://www.zora.uzh.ch/id/eprint/204774/ (Accessed: January 12, 2023).
6. Blevis, E. (2007) "Sustainable interaction design," *Proceedings of the SIGCHI Conference on Human Factors in Computing Systems*. pp. 503-512. https://doi.org/10.1145/1240624.1240705.
7. Boano, C.A. (2021) "Enabling support of legacy devices for a more sustainable internet of things," *Proceedings of the Conference on Information Technology for Social Good* [Preprint]. https://doi.org/10.1145/3462203.3475883.
8. Malinauskaite, J. and Erdem, F.B. (2021) "Planned obsolescence in the context of a holistic legal sphere and the circular economy," *Oxford Journal of Legal Studies*, 41(3), pp. 719–749. https://doi.org/10.1093/ojls/gqaa061.
9. Chirumamilla, P. (2014). The unused and the unusable: repair, rejection, and obsolescence. In Refusing, Limiting, Departing, In: CHI 2014 Workshop Considering Why We Should Study Technology Non-Use, Toronto. http://nonuse. jedbrubaker.com/wp-content/uploads/2014/03/2014_position_paper.pdf.
10. Forti, V. *et al.* (2020) *The global E-waste monitor 2020: Quantities, flows and the circular economy potential*, *UNU Collections*. United Nations University/United Nations Institute for Training and Research, International Telecommunication Union, and International Solid Waste Association. Available at: https://collections.unu.edu/view/UNU:7737 (Accessed: January 12, 2023).
11. Goodwin, C. (2020) "'Why Sideload?' user behaviours, interactions and accessibility issues around mobile app installation," *Electronic Workshops in Computing*. https://doi.org/10.14236/ewic/hci20dc.5.
12. Goodwin, C. and Woolley, S. (2022) "Barriers to device longevity and reuse: An analysis of application download, installation and functionality on a vintage device," *Reuse and Software Quality*, 13297, pp. 138–145. https://doi.org/10.1007/978-3-031-08129-3_9.
13. Gravier, M.J. and Swartz, S.M. (2009) "The Dark Side of Innovation: Exploring obsolescence and supply chain evolution for sustainment-dominated systems," *The Journal of High Technology Management Research*, 20(2), pp. 87–102. Available at: https://doi.org/10.1016/j.hitech.2009.09.001.
14. Hansson, L.Å., Cerratto Pargman, T. and Pargman, D.S. (2021) "A decade of sustainable HCI," *Proceedings of the 2021 CHI Conference on Human Factors in Computing Systems*. https://doi.org/10.1145/3411764.3445069.
15. Higginbotham, S. (2018) "The internet of trash [internet of everything]," *IEEE Spectrum*, 55(6), pp. 17–17. Available at: https://doi.org/10.1109/mspec.2018.8362218.
16. Interaction Design Foundation (2017) *What is usefulness?*, *The Interaction Design Foundation*. Available at: https://www.interaction-design.org/literature/topics/usefulness (Accessed: January 11, 2023).
17. Jensen, P.B., Laursen, L.N. and Haase, L.M. (2021) "Barriers to product longevity: A review of Business, Product Development and User Perspectives," *Journal of Cleaner Production*, 313, p. 127951. https://doi.org/10.1016/j.jclepro.2021.127951.
18. Jobin, M. *et al.* (2021) *Extending the lifetime of mobile devices to reduce their environmental impact: A glimpse on the project lifesaving*, *SocietyByte*. Marilou Jobin https://www.societybyte.swiss/wp-content/uploads/2022/02/SocietyByte_Logo_EN_plus_1030x94.png. Available at: https://www.societybyte.swiss/en/2020/05/20/extending-the-lifetime-of-mobile-devices-to-reduce-their-environmental-impact-a-glimpse-on-the-project-lifesaving/ (Accessed: January 12, 2023).

19. Kuppelwieser, V.G. *et al.* (2019) "Consumer responses to planned obsolescence," *Journal of Retailing and Consumer Services*, 47, pp. 157–165. https://doi.org/10.1016/j.jretconser.2018.11.014.
20. Makov, T. and Fitzpatrick, C. (2021) "Is repairability enough? big data insights into smartphone obsolescence and consumer interest in repair," *Journal of Cleaner Production*, 313, p. 127561. https://doi.org/10.1016/j.jclepro.2021.127561.
21. Malinauskaite, J. and Erdem, F.B. (2021) "Planned obsolescence in the context of a holistic legal sphere and the circular economy," *Oxford Journal of Legal Studies*, 41(3), pp. 719–749. https://doi.org/10.1093/ojls/gqaa061.
22. Mellal, M.A. (2020) "Obsolescence – a review of the literature," *Technology in Society*, 63, p. 101347. https://doi.org/10.1016/j.techsoc.2020.101347.
23. MichelPM (2020) *Downloading older IOS 9.3.5 versions of apps only available now for IOS 10 or later*, *Downloading Older iOS 9.3.5 Versions of A... - Apple Community*. Available at: https://discussions.apple.com/docs/DOC-13282 (Accessed: January 11, 2023).
24. Møller, K.P., Frydkjær, N.S. and Haase, L.M. (2021) *Smartphone updates as a longevity barrier for Electronic Consumer Products*, *Aalborg University's Research Portal*. LUT Scientific and Expertise Publications. Available at: https://vbn.aau.dk/en/publications/smartphone-updates-as-a-longevity-barrier-for-electronic-consumer (Accessed: January 12, 2023).
25. Nielsen, J. (2017) *Usefulness, utility, usability: 3 goals of UX Design (Jakob Nielsen)*, *YouTube*. YouTube. Available at: https://www.youtube.com/watch?v=VwgZtqTQzg8 (Accessed: January 11, 2023).
26. Proske, M. *et al.* (2016) "Obsolescence of electronics - the example of smartphones," *2016 Electronics Goes Green 2016+ (EGG)* [Preprint]. https://doi.org/10.1109/egg.2016.7829852.
27. Remy, C. (2015) "Addressing obsolescence of consumer electronics through Sustainable Interaction Design," *Proceedings of the 33rd Annual ACM Conference Extended Abstracts on Human Factors in Computing Systems*. https://doi.org/10.1145/2702613.2702621.
28. Rogerson, J. (2022) *What to do when your old iPhone or iPad doesn't run iOS 14 or iPadOS 14*, *Techradar*. Available at: https://www.techradar.com/how-to/what-to-do-when-your-old-iphone-or-ipad-doesnt-run-ios-14-or-ipados-14 (Accessed: January 11, 2023).
29. Silberman, M.S. *et al.* (2014) "Next steps for sustainable HCI," *Interactions*, 21(5), pp. 66–69. https://doi.org/10.1145/2651820.
30. Strausz, R. (2009) "Planned obsolescence as an incentive device for unobservable quality," *The Economic Journal*, 119(540), pp. 1405–1421. https://doi.org/10.1111/j.1468-0297.2009.02290.x.
31. Taffel, S., 2023. AirPods and the earth: Digital technologies, planned obsolescence and the Capitalocene. *Environment and Planning E: Nature and Space*, 6(1), pp.433-454.
32. The Economist (2009) *Planned obsolescence*, *The Economist*. The Economist Newspaper. Available at: https://www.economist.com/news/2009/03/23/planned-obsolescence (Accessed: January 11, 2023).
33. W3 (2015) *Graceful degradation versus progressive enhancement*, *Graceful degradation versus progressive enhancement - W3C Wiki*. Available at: https://www.w3.org/wiki/Graceful_degradation_versus_progressive_enhancement (Accessed: January 11, 2023).
34. Wiche, P., Pequeño, F. and Granato, D. (2022) "Life cycle analysis of a refurbished smartphone in Chile," *E3S Web of Conferences*, 349. https://doi.org/10.1051/e3sconf/202234901011.